\documentclass[useAMS]{mn2e}

\usepackage{graphicx}                                                                                                                                         
\usepackage{amssymb}

\title[HD~108: the low state at ultraviolet wavelengths]{{\em HST/STIS} spectroscopy of the magnetic Of?p star HD108: \\the low state at ultraviolet wavelengths}

\author[W. L. F. Marcolino et al.]
{\parbox{\textwidth}{W. L. F. Marcolino$^{1}$\thanks{E-Mail: wagner@astro.ufrj.br}
J.-C. Bouret$^{2,3}$, 
N. R. Walborn$^{4}$, 
I. D. Howarth$^{5}$,
Y. Naz\'e$^{6}$, 
A. W. Fullerton$^{4}$,
G. A. Wade$^{7}$, 
D. J. Hillier$^{8}$ and  
A. Herrero$^{9,10}$}\vspace{0.4cm}\\
\parbox{\textwidth}{$^{1}$Observat\'orio do Valongo, Universidade Federal do Rio de Janeiro, Ladeira Pedro Ant\^onio, 43, CEP 20080-090, Brasil \\
$^{2}$LAM-UMR6110, CNRS \& Universit\'e Provence, rue Fr\'ederic Joliot-Curie, F-13388 Marseille Cedex 13, France    \\
$^{3}$NASA/GSFC, Code 665, Greenbelt, MD 20771, USA\\
$^{4}$Space Telescope Science Institute, 3700 San Martin Drive, Baltimore, MD 21218, USA\\
$^{5}$Department of Physics and Astronomy, University College London, Gower Street, London, WC1E 6BT, UK\\
$^{6}$FNRS-GAPHE, D\'epartment AGO, Universit\'e de Liege, All\'ee du 6 Ao\^ut 17, Bat. B5C, B400-Li\`ege, Belgium\\
$^{7}$Department of Physics, Royal Military College of Canada, PO Box 17000, Station Forces, Kingston, ON K7K 7B4, Canada\\
$^{8}$Department of Physics and Astronomy, University of Pittsburgh, Pittsburgh, PA 15260  \\
$^{9}$Instituto de Astrof\'isica de Canarias, C/ V\'ia L\'actea s/n, E-38200 La Laguna, Tenerife, Spain\\
$^{10}$Departamento de Astrof\'isica, Universidad de La Laguna, Avda. Astrof\'isico Francisco S\'anchez s/n, E-38071 La Laguna, Tenerife, Spain\\ 
}}

\begin{document}   

\date{Received; Accepted }

\pagerange{\pageref{firstpage}--\pageref{lastpage}} \pubyear{2011}

\maketitle
\label{firstpage}

\begin{abstract}
  We present the first ultraviolet spectrum of the peculiar, magnetic
  Of?p star HD~108 obtained in its spectroscopic low state.  The new
  data, obtained with the Space Telescope Imaging Spectrograph ({\em STIS})
  on the {\em Hubble Space Telescope,} reveal significant changes
  compared to {\em IUE} spectra obtained in the high state:
  N$\;${\sc v}~$\lambda$1240, Si$\;${\sc iv}~$\lambda$1400, and C$\;${\sc
    iv}~$\lambda$1550 present weaker P-Cygni profiles (less
  absorption) in the new data, while N$\;${\sc iv}~$\lambda$1718
  absorption is deeper, without the clear wind signature evident in
  the high state. Such changes contrast with those found in other
  magnetic massive stars, where {\em more} absorption is observed in
  the resonance doublets when the sightline is close to the plane of
  the magnetic equator.  The new data show also that the photospheric
  Fe$\;${\sc iv} forest, at $\sim$1600--1700\AA, has strengthened
  compared to previous observations.  The ultraviolet
  variability is large compared to that found in typical, non-magnetic
  O stars, but moderate when compared to the high-/low-state
  changes reported in the optical spectrum of HD~108 over several
  decades.  We use non-LTE expanding-atmosphere models to analyze the
  new {\em STIS} observations. Overall, the results are in accord with a
  scenario in which the optical variability is mainly produced by
  magnetically constrained gas, close to the photosphere. The
  relatively modest changes found in the main ultraviolet wind lines
  suggest that the stellar wind is not substantially variable on a
  global scale. Nonetheless, multidimensional radiative-transfer
  models may be needed to understand some of the phenomena observed.
\end{abstract}



\maketitle

\section{Introduction}

The Of?p spectroscopic classification was introduced to describe a 
small group of O-type stars showing emission in the C$\;${\sc iii}
lines at 4650\,\AA\ at a strength comparable to that of the N$\;${\sc
  iii} lines at 4634--42\,{\AA} (Walborn 1972). Normally, C$\;${\sc iii} is
  weaker, or absent, in Of-star spectra. To date,
only five stars belonging to the Of?p class have been found in the
Galaxy: HD~108, HD~148937, HD~191612, NGC~1624-2, and
CPD$-28^{\circ}2561$, with the last two identified only recently
(Walborn et al.\ 2010).

Substantial spectroscopic variability is now known to be an inherent
property of the Of?p stars, with hydrogen and helium lines in the
optical observed to vary between emission and absorption states on
timescales that range from days to decades in different objects
(Naz\'e et al. 2001; Walborn et al. 2003); the C$\;${\sc
  iii}~$\lambda$4650 emission varies in concert with the hydrogen
lines.  The spectroscopic changes appear to correlate with low-level
photometric variability, such that the stars are a few per cent
brighter in the optical when H$\alpha$ has maximum emission (e.g.,
Howarth et al. 2007; Barannikov 2007).  Throughout this paper, the
optical-spectrum emission and absorption states will be referred to as
`high' and `low' states.
 
\begin{figure}
\includegraphics[width=7cm,height=9cm,trim= 0mm 0mm 0mm 25mm,clip,angle=-270]{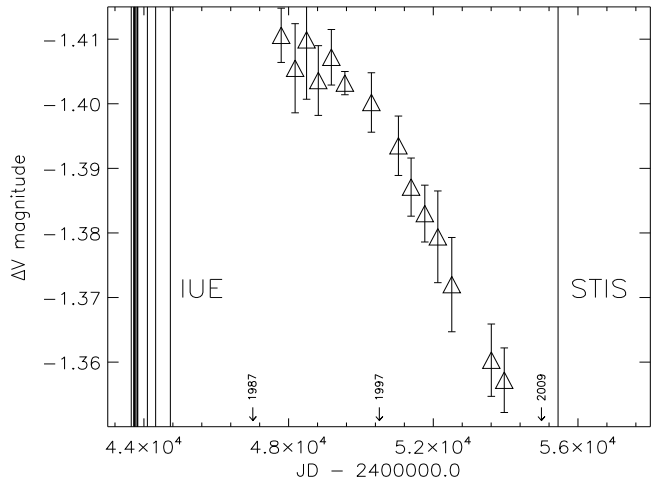}
\caption{V-magnitude light-curve of HD~108 as a function of the Julian Day 
(data from Barannikov 2007). The observation dates of the {\em IUE} (made before 1982) 
and {\em STIS} (2010) data are indicated by vertical lines; photometric maximum corresponds to the 
spectroscopic high state ({\em IUE} observations), and vice versa. 
Arrows indicate observation dates of spectra displayed in Fig. \ref{optical}.}
\label{photometry}
\end{figure}

A crucial step towards an understanding of the Of?p phenomenon was
made recently, with the detection of surface magnetic fields in
HD~191612, HD~148937, and HD~108 (Donati et al.\ 2002; Wade et al.\ 2011; 
Hubrig et al.\ 2008, 2011; Martins et al.\ 2010 - hereinafter Paper I).
The spectral variability is associated with magnetically constrained gas,
whose aspect is rotationally modulated. The presence of a magnetic
field could also potentially explain the exceptional X-ray
luminosities of Of?p stars, observed to be above the O-star average
(Naz\'e et al. 2010).

Ultraviolet (UV) spectra of massive stars are of particular interest,
not least because they present the most sensitive diagnostics of
stellar winds; but while the optical spectra of Of?p stars have been
studied in detail for several years (e.g., Naz\'e et al.\ 2001, 2010),
characteristics of their ultraviolet spectra are poorly known.
For example, it is unclear to what extent the UV is variable (if at
all), despite some efforts based on the sparse available data (e.g.,
Howarth et al.\ 2007). Are line-profile changes as intense as those 
observed in the optical?  What are the consequences in terms of
stellar physical properties?

Until now, we have lacked UV spectra for Of?p stars taken at
contrasting optical states but with consistent wavelength intervals and
comparable resolutions.  Consequently, it is not yet known whether
their wind properties are subject to change (either globally, or
simply as a viewing effect), or to what extent the magnetic field
influences the stellar atmosphere as a whole. In this context, new
observational data are crucial for a better understanding of the Of?p
class and the effects of magnetic wind confinement.  They can also
provide empirical information to be compared to predictions provided
by MHD simulations (e.g., ud-Doula et al. 2009)

\begin{figure}
\includegraphics[width=9cm,height=6.5cm]{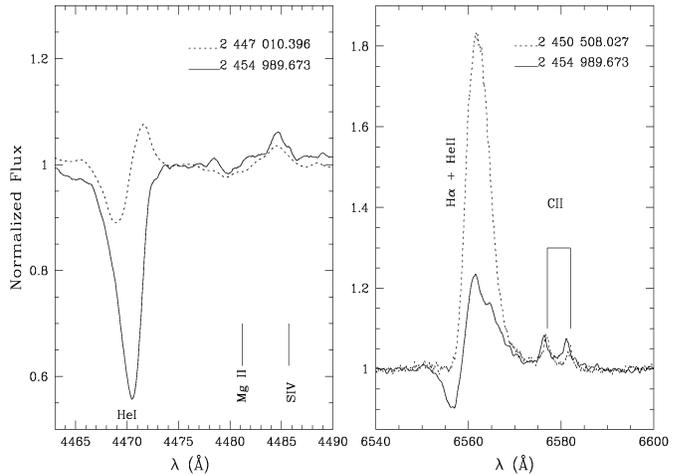}
\caption{The low state of HD~108 (solid line; data from 2009) and two higher states 
(dotted-lines; data from 1987 and 1997). Exact Julian Dates are indicated. 
Note the drastic profile changes of He$\;${\sc i}~$\lambda$4471  
and H$\alpha$. These states correlate well with photometric data 
(see arrows in Fig. \ref{photometry}).}
\label{optical}
\end{figure}

In the present paper, we report the first high-resolution ultraviolet
spectra of the Of?p star HD~108 in its low state. HD~108 is hot
($\sim$ 35 kK), massive ($\sim$ 40 M$_\odot$), and harbours a relatively
intense magnetic field (1--2~kG; Paper I).  This field is
considered to be responsible for the observed optical variability,
modulated on a rotational period of about 55 years. Since HD~108 can
be considered a prototype of the Of?p class (Walborn 1972), we expect
the results and conclusions drawn from the present work to be
applicable to other Of?p stars.

The paper is structured as follows: in Section \ref{uvsection} we
present and describe the new UV data, acquired with the Hubble Space
Telescope ({\em HST}). We compare them with archival spectra from the
International Ultraviolet Explorer ({\em IUE}) satellite, obtained
when HD~108 was in its optical high state. The main differences are
discussed.  Section \ref{models} presents expanding-atmosphere
(CMFGEN) models used to analyze the observations.  Possible changes in
the physical parameters with respect to previous analyses are discussed.
The main results and conclusions are summarized in Section
\ref{discussion}, where the significance of ultraviolet variability of
HD~108, and the consequences for the interpretation of the Of?p
phenomenon, are reviewed.

\begin{figure*}       
\centering       
\includegraphics[width=14.5cm,height=18.8cm,angle=-270]{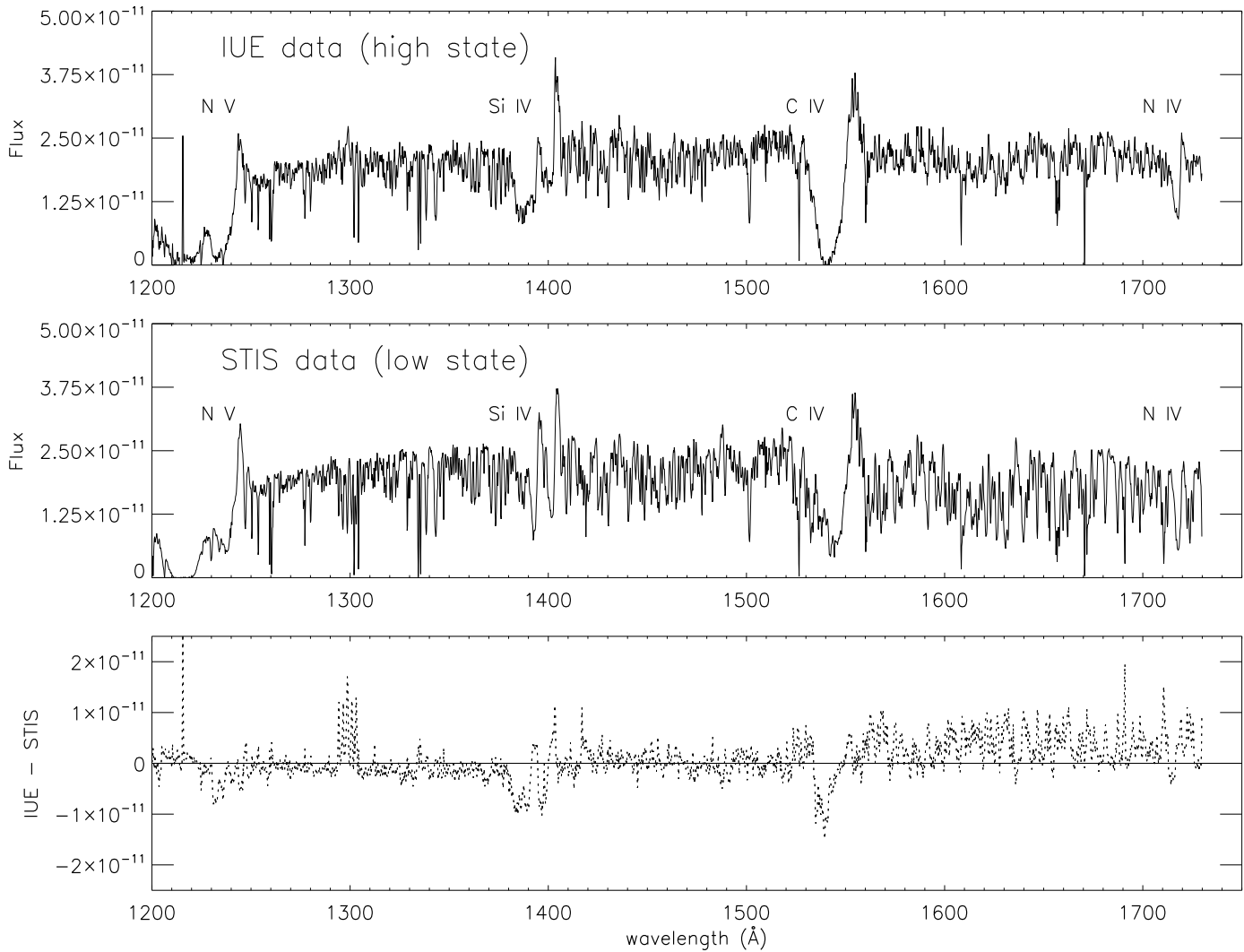}       
\caption{Ultraviolet spectra of HD 108 in the 1200-1750\AA\, range. Upper panel: {\em IUE} 
data acquired at the high state (SWP08352). Middle panel: new {\em STIS}
data in the low state. Bottom panel: difference spectrum. Flux units are erg cm$^{-2}$ s$^{-1}$ \AA$^{-1}$.}      
\label{uvplot}       
\end{figure*}  

\section{ULTRAVIOLET SPECTRoscopy}
\label{uvsection}

We used the Space Telescope Imaging Spectrograph ({\em STIS}) on board {\em
HST} to obtain ultraviolet spectra of HD~108 in 2010 September
(P.I. Bouret).  The observations were made using the FUV-MAMA and
NUV-MAMA detectors, with the E140M and E230M echelle gratings. The
wavelength coverage is $\sim$1150--1750 \AA\ (E140M) and 2250--3150 \AA\
(E230M), with resolving powers of $R = 45,800$ and 30,000,
respectively.  The signal-to-noise ratio (SNR) of the data is $\sim$30.
By reference to results reported by Naz\'e et al.\ (2010) and 
photometry illustrated in Fig.~\ref{photometry}, we infer that the 
observations were taken during the lowest emission-line state reached 
by HD~108, when it is faint in the BVR bands (Barannikov 2007) and the
optical hydrogen and helium lines are in absorption, or have minimum
emission compared to the high state. As an example, Fig.~\ref{optical} 
presents optical profiles of He$\;${\sc i}~$\lambda$4471 and H$\alpha$ 
observed recently (in 2009) and decades ago (in 1997 and 1987).

Figure \ref{uvplot} presents the flux-calibrated {\em STIS} data for
1200--1750 \AA, where the main wind lines occur.  For comparison, we
also plot {\em IUE} data (SWP08352) obtained when HD~108 was at its
optical high state.  Other {\em IUE} SWP 
observations show essentially the same spectrum, since they 
were also taken in the high state (Fig.~\ref{photometry}).
The resolving power and SNR of the {\em IUE} 
data ($R \simeq 10^4$ and $\sim$10, respectively) are poorer 
than for {\em STIS}, so both datasets were convolved to an 
effective resolution of 0.2 \AA\ (FWHM).

\begin{figure*}
\centering
\includegraphics[width=10cm,height=14cm,angle=-270]{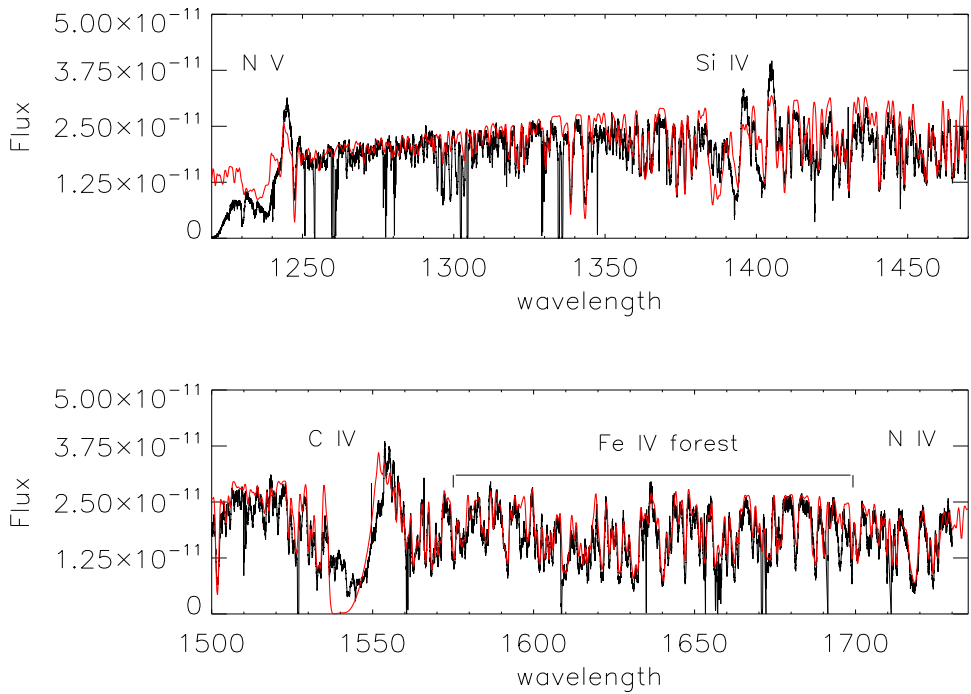}
\caption{{\em STIS} low-state spectrum of HD~108 (black) compared with the
  adopted CMFGEN model (red).  The stellar and wind parameters are as
  reported in Paper I, with the exception of the
  microturbulent velocity (Section~\ref{turb}; see Table \ref{params}). Note that, in this state, C$\;${\sc
    iv}~$\lambda$1550 is observed to have less absorption than
  predicted by the model
  (see text for more details). Flux units are erg cm$^{-2}$ s$^{-1}$
  \AA$^{-1}$.}
\label{finalmodelfit}
\end{figure*}

\subsection{Variability}

Two general conclusions can immediately be drawn from
Fig.~\ref{uvplot}. First, we do not observe {\em drastic} changes in
the UV spectrum of HD~108 -- that is, not as drastic as those observed
in the optical spectrum, where intense high-state emission lines have
turned into absorption lines over recent decades (Fig. \ref{optical}; see also Naz\'e et al.\
2001; 2010).  The continuum level remains constant to within
the combined observational uncertainties ($\sim$5\%),
so that the Of?p phenomenon does not cause the energy flux
distribution in the UV to vary substantially.  Moreover, the same basic set of
photospheric and wind lines is present in both states, although with
somewhat different profiles (see below).  Qualitatively, it therefore
seems that the global structure of the stellar wind is not
severely modified by the magnetic field and that, with appropriate
caveats, we may reasonably attempt to model the data with spherically symmetric
models in the first instance.

On the other hand, it is clear that the observed variability very
considerably exceeds that found in the UV spectra of {\em normal} O
stars (cf., e.g., Kaper et al.\ 1996).  There are clear changes
between high- and low-state spectra in all the major wind lines:
N$\;${\sc v}~$\lambda$1240, Si$\;${\sc iv}~$\lambda$1400, C$\;${\sc iv}
$\lambda$1550, and N$\;${\sc iv}~$\lambda$1718 (see the difference
spectrum in Figure \ref{uvplot}).  In the low-state {\em STIS} spectrum, the
subordinate N$\;${\sc iv}~$\lambda$1718 line has a deep/broad absorption
profile without red wind emission, while the resonance lines exhibit
P-Cygni profiles with reduced absorption. In particular, C$\;${\sc iv}
$\lambda$1550 now presents an unsaturated profile.  The strengths of
the P-Cygni emission components do not change, however, with the exception of
N$\;${\sc v}~$\lambda$1240, which is somewhat stronger in the {\em STIS} spectrum 
than in the {\em IUE} data.  As it is discussed later in the paper 
(see Section \ref{discussion}), a quantitative explanation for this fact, 
as well as for the decrease in the absorption of the resonance lines awaits 
more sophisticated MHD simulations. Nevertheless, we note that 
the different behavior of N$\;${\sc v}~$\lambda$1240 may be related 
to its higher sensitivity to X-rays (Hillier \& Miller 2003; Marcolino et al. 2009). 
A small change in the number of X-rays photons available could potentially vary 
its emission level via Auger ionization. The photospheric iron forest also 
exhibits changes (especially beyond 1500\AA), being considerably stronger 
in the {\em STIS} data.

{\em STIS} data at wavelengths larger than 2250 \AA\ (not shown) are comparatively featureless,
with few, weak photospheric lines. An exception is C$\;${\sc iii}
$\lambda$2296, which presents a relatively deep absorption. A detailed
comparison with {\em IUE} spectra obtained at the high state is
hindered by their resolution and, particularly, very poor SNR in this
region.

\subsection{Comparison with related systems}

Interestingly, the changes in the P-Cygni profiles observed for HD~108
do not correlate with rotational phase in the same way as reported for
magnetic B stars (see for example Shore et al. 1990; Henrichs et al. 2005). 
These stars normally present more absorption --
e.g., in C$\;${\sc iv} -- when the observer's line of sight lies close to
the plane of the magnetic equator. Such configuration corresponds 
to the low state of Of?p stars (see Wade et al. 2011b), i.e., 
the one analyzed in this paper. 

The usual interpretation for magnetic 
B stars is that gas constrained by the magnetic field settles in a 
cooling disk about the magnetic equator; this disk would provide 
the extra absorption when observed edge-on (see however Section \ref{discussion}).
We note however that such absorptions generally occur near 
the line center (i.e., low velocities), which points to corotating 
disk-like structures rather than a rapidly expanding wind.


The magnetic O dwarf star $\theta^1$ Ori C follows the enhanced absorption 
scenario (Smith \& Fullerton 2005). When its magnetic equator is edge-on ($\phi \simeq 0.5$) C$\;${\sc
  iv}~$\lambda$1550 shows enhanced blueshifted absorption relative to
$\phi \simeq 0$.  The red side of this feature also presents a small
increase in flux at $\phi \simeq 0.5$; that is, the P-Cygni profile as
a whole seems stronger when the magnetic equator is edge-on (see
Fig.~3 in Smith \& Fullerton 2005).  At about this same phase, the
H$\alpha$ profile has its minimum emission (Stahl et al.\ 1996).  However,
the situation in HD~108 is different: less absorption in C$\;${\sc iv}
$\lambda$1550 {\it and} minimum emission in H$\alpha$ are observed when the
line of sight lies close to the plane of the magnetic equator.  This
issue is discussed further in Section \ref{discussion}.

Besides HD~108 and $\theta^1$ Ori C, four further O~stars have
detected magnetic fields: HD~191612, HD~148937, HD~57682, and $\zeta$
Ori A (Donati et al.\ 2006; Hubrig et al. 2008; Wade et al.\ 2011; Grunhut et al.\ 2009;
Bouret et al.\ 2008). While all have estimated rotation periods, available
UV data are insufficient to draw conclusions concerning wind-profile
behaviour at specific phases. An exception is HD~191612. New 
UV observations of this star were recently acquired by our group and a 
detailed analysis will be presented in a forthcoming paper.

\section{Model atmospheres}
\label{models}

We have used CMFGEN (Hillier \& Miller 1998) to generate non-LTE
models to analyze the new {\em STIS} observations of HD~108.  CMFGEN solves
the equation of radiative transfer in a spherically symmetric outflow
under the constraints of radiative and statistical equilibria.
Further model assumptions, and basic procedures for the determination
of parameters, are as described in detail in Paper~I.

In Paper~I we obtained a reasonable fit to optical (low state), UV ({\em IUE};
high state), and far-UV ({\em FUSE}; low state) data of HD~108 with a
single model. Since the UV and far-UV observations were obtained in
different states, it was tentatively concluded that the global
stellar-wind properties may be stable in spite of the large changes
occuring in the optical lines (and attributed to a localised,
magnetically constrained plasma). In order to examine this idea, we
began the spectral analysis with the set of parameters derived in
Paper~I.  After several exploratory models, we concluded that,
notwithstanding some discrepancies, these basic parameters also
represent the new {\em STIS} data reasonably well. The problems and
improvements found during our analysis are discussed below. Our
adopted best-fit model is shown in Fig.~\ref{finalmodelfit} and 
the stellar and wind properties are summarized in Table \ref{params}.

\subsection{Mass-loss rate}

In order to desaturate the model C$\;${\sc iv} $\lambda$1550 P-Cygni profile,
to better match the {\em STIS} data (Fig.~\ref{uvplot}), we initially
adjusted  the mass-loss rate $\dot{M}$ to decrease the
stellar-wind density.  A match could only be achieved by reducing
the mass-loss rate by an order of magnitude (to $\sim$10$^{-8}$
M$_\odot$ yr$^{-1}$). However, while the observed intensities of C$\;${\sc iv}
$\lambda$1550 and N$\;${\sc v}~$\lambda$1240 are then correctly
predicted, their full profiles are not reproduced in detail -- the
synthetic lines lack P-Cygni absorption at low velocities.
Furthermore, the C$\;${\sc iii}~$\lambda$1176 and Si$\;${\sc
  iv}~$\lambda$1400 lines are predicted to be purely photospheric,
while {\em STIS} data reveal conspicuous P-Cygni profiles 
(see Section~\ref{turb} for fits to C$\;${\sc iii}~$\lambda$1176).

We could find no model that simultaneously reproduces, in detail, both
C$\;${\sc iv}~$\lambda$1550 and the weaker P-Cygni profiles observed in
the {\em STIS} spectrum (C$\;${\sc iii}~$\lambda$1176, N$\;${\sc v}~$\lambda$1240,
Si$\;${\sc iv}~$\lambda$1400). We explored other ways to decrease the 
intensity of the wind profiles (e.g., adjusting the wind clumping 
and the X-ray distribution and luminosity) but they all turned out to be unsuccessful.
Nevertheless, given the problems that arose in adopting lower mass-loss rates, we consider the
value of $\dot{M} = 10^{-7}$ M$_\odot$ yr$^{-1}$ obtained in Paper~I
to be reasonable.  However, in the light of the spectroscopic
variability, we now adopt a more conservative uncertainty of $\pm$1~dex for
this parameter. We return to this issue in Section~\ref{discussion}.

\begin{table}
\caption{Stellar and wind properties of HD108.}
\begin{tabular}{ll}
\hline
T $_{eff}$ (K)           & 35 000 $\pm$ 2000 \\
log $g$ (cgs)            & 3.50 $\pm$ 0.20     \\
log L$_\star$/L$_\odot$   & 5.70 $\pm$ 0.10    \\
R$_{\star}$/R$_\odot$     & 19.2 $^{+3.3}_{-2.8}$ \\
M$_{\star}$/M$_{\odot}$    & 43  $^{+32}_{-18}$    \\
$v_{\rm eq}\sin{i}$   (km s$^{-1}$) & 0  \\
$v_{mac}$ (km s$^{-1}$)   & 45  \\
$\xi _t^{phot}$  (km s$^{-1}$)  & 50  \\
P$_{rot}$ (yr)           & 50-60 \\
\hline
log $\dot{M}$            & -7.0 $\pm$ 1.0  \\
v$_\infty$ (km s$^{-1}$)   & 2000 $\pm$ 300 \\
log $L_X/L_{BOL}$        & -6.2 \\
\hline
\end{tabular}
\label{params}
\end{table}

\subsection{Effective temperature and Fe abundance}

The new {\em STIS} data show that the numerous Fe$\;${\sc iv} lines 
in the $\sim$1600--1700 \AA\ region -- the iron forest --
are stronger than in
the high-state observations (Fig. \ref{uvplot}).  A possible cause for
this variation is a change in the effective temperature, $T_{\rm
  eff}$, which affects the balance between Fe$\;${\sc iii}, Fe$\;${\sc
  iv}, and/or Fe$\;${\sc v} line intensities in the UV spectra of OB
stars (the influence of other parameters is
examined in the Section~\ref{turb}).  We explored models with
effective temperatures both
higher and lower than the baseline value of 35 kK. The
more intense iron forest in the {\em STIS} spectrum is best reproduced with
a slightly reduced $T_{\rm eff}$, in the range 32.5--34 kK. However,
for the lower value of $T_{\rm eff}$ both C$\;${\sc iii}~$\lambda$1176 and Si$\;${\sc
  iv}~$\lambda$1400 have very strong wind signatures in the models, in
stark contrast with the observations.

Our conclusion is that the new {\em STIS} data are better matched with a
slightly lower effective temperature than the value inferred in Paper~I, which
was based only on helium lines. However, the difference is comfortably
within the previously adopted uncertainty of 2 kK.

Given the results reported above and in the previous section regarding
the C$\;${\sc iii}~$\lambda$1176 and Si$\;${\sc iv}~$\lambda$1400 lines, we
also explored the effects of models in which effective temperatures
below 35 kK were coupled
with lower mass-loss rates. These models were unsuccessful in
fitting all wind lines simultaneously; as the mass-loss rate
decreases, the initially strong C$\;${\sc iii}~$\lambda$1176 and Si$\;${\sc
  iv}~$\lambda$1400 features go into absorption long before C$\;${\sc
  iv}~$\lambda$1550 becomes desaturated.

Our models indicate that changes in the iron content also modify the 
strength of the Fe$\;${\sc iv} forest, as expected. An 
inhomogeneous distribution of iron in the photosphere, for example, could 
cause variability according to the stellar rotational period. 
However, we think this scenario is unlikely. It is hard to justify 
an increase in the abundance of this element much above the solar 
value (needed to fit the observations). Moreover, we recall that 
other lines are also stronger in the {\em STIS} data 
(compared to {\em IUE}), e.g., N$\;${\sc iv}~$\lambda$1718 and 
He$\;${\sc ii}~$\lambda$1640. In the next section we discuss the parameter 
that was able to improve the fits to all these transitions and also 
the {\em FUSE} region, simultaneously.

\begin{figure}
\centering
\includegraphics[width=5.5cm,keepaspectratio,trim= 13mm 17mm 13mm 17mm,angle=-270]{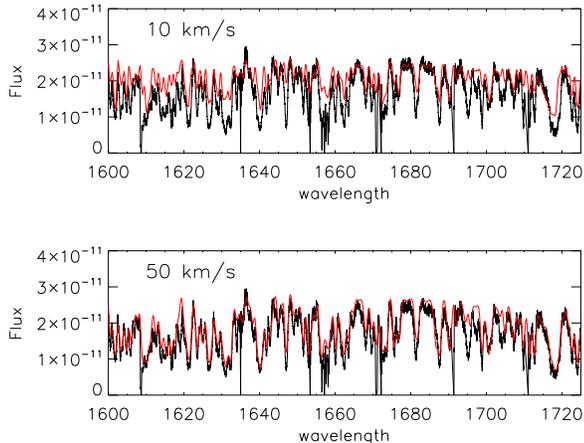}
\caption{Effect of microturbulence on the modelled spectrum. 
Top panel: {\em STIS} spectrum (black) and model with $\xi^{\rm phot}_{\rm t}$ = 10 km s$^{-1}$ (red). Lower 
panel: {\em STIS} spectrum and model with $\xi^{\rm phot}_{\rm t}$ = 50 km s$^{-1}$. 
Better agreement is achieved using the higher $\xi^{\rm phot}_{\rm t}$ for the Fe$\;${\sc iv} forest, He$\;${\sc ii}~$\lambda$1640, 
and N$\;${\sc iv}~$\lambda$1718.  Flux units are erg cm$^{-2}$ s$^{-1}$ \AA$^{-1}$.}
\label{turbstis}
\end{figure}

\begin{figure}
\centering
\includegraphics[width=5.5cm,keepaspectratio,trim=13mm 17mm 13mm 17mm,angle=-270]{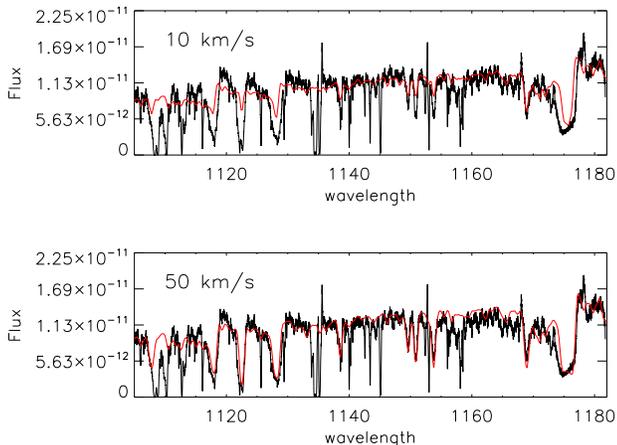}
\caption{As Fig.~\ref{turbstis} but for the {\em FUSE} spectrum.  The
  deep, narrow absorption lines that are not predicted are of
  interstellar origin. A better agreement to most lines is achieved
  with a large turbulent velocity.}
\label{turbfuse}
\end{figure}

\subsection{Rotation, macroturbulence, and microturbulence}
\label{turb}

Other parameters affect the line profiles in the Fe$\;${\sc iv}
forest, including the projected equatorial rotation velocity, $v_{\rm
  eq}\sin{i}$; the macroturbulence; and the microturbulence.  The
first and the second parameters do not alter the line equivalent
widths, $W_\lambda$, but merely redistribute the flux.
Here, following the analysis performed in Paper~I, we adopt 
a $v_{\rm eq}\sin{i}$ of zero (consistent with the inferred 
very long rotation period) and a gaussian macroturbulent velocity
of $v_{\rm mac} = 45\,\mbox{km s}^{-1}$.  Lower values produce deeper
synthetic Fe$\;${\sc iv} profiles, but they are much narrower than
observed (since $W_\lambda$ is conserved). We recall that 
the  origin, and hence a proper description, of the macroturbulence is
still a matter of debate (see Sim\'on-D\'iaz et al.\ 2010 and
references therein).

The intensity of the Fe$\;${\sc iv} forest is also affected by the
microturbulence, $\xi_{\rm t}$. In CMFGEN, the comoving frame 
calculations are performed using $\xi^{\rm phot}_{\rm t}$, while the 
spectrum in the observer's frame is computed by using a radially dependent 
turbulence parametrized by
\[ \xi_{\rm t}(r) = \xi^{\rm phot}_{\rm t} + (\xi^{\rm max}_{\rm t} - 
\xi^{\rm phot}_{\rm t})\frac{v(r)}{v_\infty}, \] 
where $v(r)$ and $v_\infty$ are the velocity field and terminal speed, respectively, 
and $\xi^{\rm max}_{\rm t}$ corresponds to the maximum turbulent velocity (assumed 
to be about 10\% of the terminal speed). This prescription mimics
classical microturbulence at the photospheric level and the velocity dispersion 
from shocks occuring in the stellar wind (see Hillier et al.\ 2003 for details). 

By varying $\xi^{\rm phot}_{\rm t}$ from an initial value
of 10 km s$^{-1}$ up to 50 km s$^{-1}$ we found progressively better
fits to the iron forest in the {\em STIS} spectrum.  Surprisingly, we also
found a substantial improvement in the fits of a number of other
lines. For example, we achieved excellent agreement with the deep,
broad N$\;${\sc iv}~$\lambda$1718 line, and could much better reproduce
the far-UV {\em FUSE} spectrum, which was analyzed in Paper~I but with
conspicuous discrepancies in line width and depth.
We recall that neither $v_{\rm eq}\sin{i}$, nor different macroturbulent values, were
able to improve the fit to all these transitions.

Figs.~\ref{turbstis} and~\ref{turbfuse} illustrate the effects of
changing $\xi^{\rm phot}_{\rm t}$. We note that {\em FUSE} and 
{\em STIS} data below 1200\AA\, are virtually identical, both 
being acquired in the low state. Although values of $\xi^{\rm
  phot}_{\rm t} \simeq 50\, \mbox{km s}^{-1}$ appear unrealistic for 
typical O~stars, our results hint that some unknown turbulence
may be at work, particularly at the configuration corresponding
to the low state (cooling disk/magnetic equator edge-on). Note that
such high turbulent velocities also affect the optical
spectrum. However, the best fit we could achieve for this spectral region
presents difficulties due to the observed presence of
core line emissions and abnormally deep absorption profiles (see
Paper~I), hampering a reliable $\xi^{\rm phot}_{\rm t}$
determination.

\subsection{Line-formation regions vs Alfv\'en radius}

In order to gain more insight into the influence of the magnetic
field on the atmosphere of HD~108 we examined the radial extents of
line-formation regions in our final model (Fig.~\ref{finalmodelfit})
and compared them to the location of the Alfv\'en radius, calculated from
\[ \left( \frac{R_{\rm A}(\theta)}{R_\star} \right)^{2q-2} - \left(
  \frac{R_{\rm A}(\theta)}{R_\star} \right)^{2q-3} =
\eta_\star(4-3\sin^2\theta) \]
(ud-Doula \& Owocki 2002).  For simplicity, we assume a confinement
parameter $\eta_\star = 100$ (Paper~I), $\theta = 90^\circ$ (line of
sight in the plane of the magnetic equator), and $q = 3$ (dipole
field), giving $R_{\rm A} \simeq 3.45\,R_\star$.


As an example, Fig.~\ref{LFRs} shows the formation regions of the
N$\;${\sc iv} $\lambda$1718, C$\;${\sc iv} $\lambda$1550, 
Si$\;${\sc iv} $\lambda$1400, H$\alpha$, and H$\beta$ lines. 
The location of the Alfv\'en radius is indicated.  
The C$\;${\sc iv} and Si$\;${\sc iv} features are formed 
in a very extended region, encompassing $R_{\rm A}$. 
N$\;${\sc iv} $\lambda$1718 is formed closer to the photosphere 
but at the same time it goes farther than the Balmer lines.
This is compatible with the moderate changes observed in the UV 
(Figure \ref{uvplot}); the magnetic field is expected to have a 
relatively modest influence as we get closer to the Alfv\'en radius and beyond. 
In contrast, H$\alpha$ and H$\beta$ are formed well below $R_{\rm A}$, 
where the influence of the magnetic field on the gas distribution is 
more significant (see ud-Doula \& Owocki 2002).

Our models, while providing a reasonable fit in the UV, predict
pure-absorption Balmer profiles at all times. 
However, observations reveal large variability of some optical lines.
For example, H$\beta$ has a very intense emission at the high 
state and an absorption profile at minimum (Naz\'e et al.\ 2010; 
see also the behavior of H$\alpha$ in Fig.~\ref{optical}). Fig.~\ref{LFRs} thus
suggests that the optical variability/emission observed is probably related to
material constrained by the magnetic field close to the photosphere, below $R_{\rm A}$.
Currently, this constrained material is not modelled by CMFGEN or any
other detailed atmosphere model.


\section{DISCUSSION AND CONCLUSIONS}
\label{discussion}

The intense optical variability presented by Of?p stars has been
described in detail over the last decade. However, until now it has
been unclear if variability would also be present at ultraviolet
wavelengths.  Consequently, it has been a matter of speculation as to
whether the global stellar wind parameters of Of?p stars change
through the different optical states observed.

A first qualitative result in this sense was reported by Howarth et
al.\ (2007), for the Of?p star HD~191612. They reported that
two UV spectra (from {\it IUE}) obtained in epochs of different optical spectra showed
no large differences, suggesting that the global wind properties of
HD~191612 (e.g., mass-loss rate) are not significantly modulated by the
Of?p phenomenon.  However, the comparison made by those authors was
limited by having to use spectra at very different spectral
resolutions, requiring degradation of the higher-resolution
observation ($\sim$0.2\AA) to match the lower-resolution data
($\sim$6\AA). As a consequence, moderate variability could not be
ruled out.  Despite this, the large optical line changes observed
were attributed to magnetically constrained material, mainly close to the photosphere, 
since otherwise the UV wind profiles would similarly be highly variable.

\begin{figure}
  \includegraphics[width=7.5cm,height=9.5cm,angle=-90]{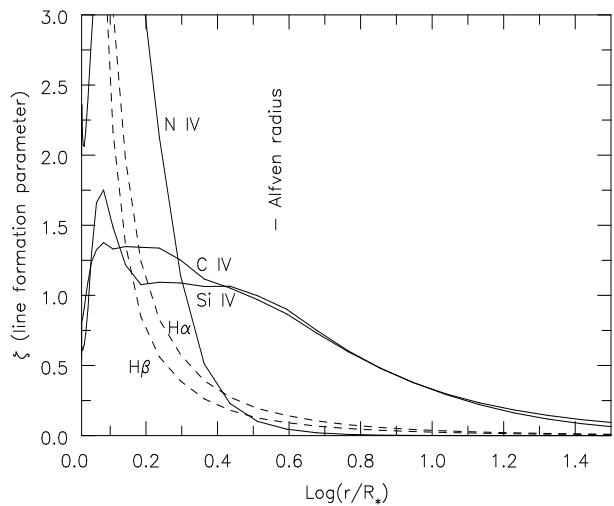}
\caption{Line-formation regions of H$\alpha$, H$\beta$, N$\;${\sc iv} $\lambda$1718, 
C$\;${\sc iv} $\lambda$1550, and Si$\;${\sc iv} $\lambda$1400. The Balmer and UV transitions 
are represented by dashed and solid lines, respectively. The location of 
the Alfv\'en radius ($R_{\rm A}$) is indicated. Note that the Balmer lines are formed well 
below $R_{\rm A}$, where the influence of the magnetic field is greatest.}
\label{LFRs}
\end{figure}

The present paper casts light on these issues by presenting, for the
first time, new high-resolution ultraviolet spectra of HD~108 in the
low state (very close to minimum; Naz\'e et al.\ 2010), alongside
high-state data of comparable quality. The main results from our
analysis are:

\begin{itemize}

\item The UV spectrum of HD~108 does not present variability as
  dramatic as that seen in the optical, where previously intense
  emission lines have changed into absorption features (or weak emission) 
  in recent decades (see Fig. \ref{optical}; see also 
  Naz\'e et al.\ 2010). On the other hand, we emphasize that the UV variability is
  very large compared to that usually found in typical,
  non-magnetic O stars (see for example Kaper et al.\ 1996).\\

\item The new {\em STIS} data reveal moderate differences compared to previous
  {\em IUE} data acquired at the high state. The low-state spectrum
  has somewhat less intense P-Cygni profiles (e.g., in N$\;${\sc
    v}~$\lambda$1240 and C$\;${\sc iv}~$\lambda$1550).  The photospheric
  iron forest is stronger, and the N$\;${\sc iv}~$\lambda$1718 line is
  deeper than in the high state,
  without an obvious wind profile. The UV continuum level remains
  constant, to within observational errors, between the low and high states.\\

\item The analysis of the UV spectra through expanding-atmosphere
  CMFGEN models suggests that the wind properties of HD~108 at the
  optical low and high states remain essentially preserved, in
  agreement with the suggestion by Howarth et al.\ (2007) in the case
  of HD~191612. However, simultaneous fits to all UV wind lines proved
  elusive, suggesting that the mass-loss rate determination is
  somewhat uncertain. More sophisticated
  models may be needed to solve this issue.\\

\end{itemize}

\begin{figure}
\includegraphics[width=9cm,height=7cm,trim= 20mm 3mm 0mm 0mm,clip]{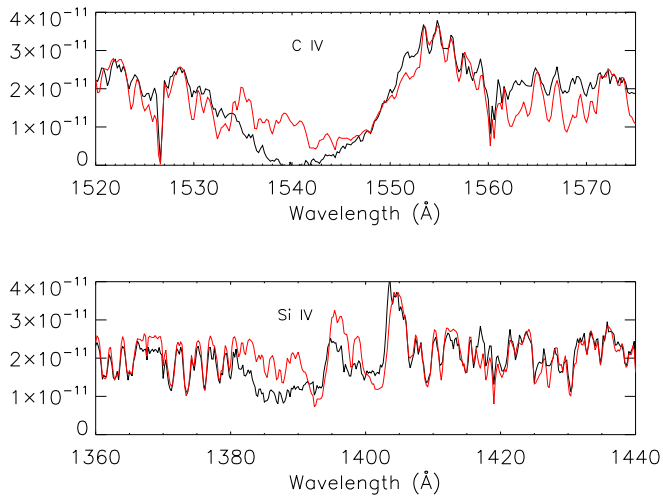}
\caption{C$\;${\sc iv}~$\lambda$1550 and Si$\;${\sc iv}~$\lambda$1400 variability in HD~108. The profiles 
present more absorption at the high state (black line; {\em IUE} data) and are weaker at the low state 
(red line; magnetic equator edge-on; {\em STIS}). Other massive magnetic stars reveal the opposite 
when the magnetic equator/cooling disk is seen edge-on. Flux units are erg cm$^{-2}$ s$^{-1}$ \AA$^{-1}$.}
\label{civ}
\end{figure}

Although moderate in scale compared to the optical, the variability 
revealed by the {\em HST} data carries important new information. The 
P-Cygni profiles of HD~108 vary in an opposite sense to those found in
other magnetic OB stars. Usually, when the magnetic equator/cooling
disk is seen edge-on, more absorption is detected in the UV resonance
lines (e.g., $\theta^1$ Ori C; Smith \& Fullerton 2005). In HD~108, in
this configuration, we detect {\em less} absorption in the wind
profiles. This is illustrated for C$\;${\sc iv}~$\lambda$1550 and Si$\;${\sc
  iv}~$\lambda$1400 in Fig.~\ref{civ}. 

At a first glance, more absorption is compatible with 
the density enhancements expected in the magnetic equatorial 
region due to gas confinement.  However, this constrained 
material has also lower velocities compared to the flow at 
higher latitudes (e.g., ud-Doula et al.\ 2002). Therefore, the absorption 
should be less blueshifted and is expected to be less intense as a 
whole, as reported here for HD~108 (Sundqvist \& ud-Doula; private comm.).
In fact, MHD simulations for $\theta^1$ Ori C fail to describe the 
observed C$\;${\sc iv}~$\lambda$1550 variability, predicting a less intense 
absorption profile when $\phi \sim 0.5$ (ud-Doula 2008). 

An explanation for the contrasting behavior of the UV wind line 
variability in HD~108 and $\theta^1$ Ori C (and other magnetic B stars) may reside in the nature 
of their cooling disks. Host stars with different physical parameters and 
magnetic field strengths could possess disks with different properties 
(e.g., ionization, density, and temperature). More or less absorption 
could be dictated by different ionization/recombination rates in 
each case. Indeed, HD~108 has a much higher mass ($\sim 40\,M_\odot$) 
and luminosity ($\sim 5\times{10^5}L_\odot$) than do B stars. 
Its luminosity, temperature, and radius are also considerably different 
from $\theta^1$ Ori C (e.g., Donati et al. 2002). New MHD simulations 
coupled with radiative transfer calculations are needed to address such questions.

An alternative hypothesis for the UV variability in HD~108 which 
does not involve material accumulated in the magnetic equatorial region 
is simply a time-dependent stellar wind (stochastic variability). In
this case, we would expect no correlation between rotational phase and
the UV and optical line changes.  Further monitoring would test this
scenario.

For more constraints on these questions, it would be of interest to
investigate whether the effect observed in Fig.~\ref{civ} occurs in
other Of?p stars. HD~191612 is particularly important, since its
magnetic and rotational properties are relatively well studied
(Howarth et al.\ 2007, Wade et al.\ 2011b).  In this star, the
H$\alpha$ equivalent-width and longitudinal-field measurements
($B_{\rm Z}$) are very reproducible on the 537.6-d rotation period
(see Fig. $3$ in Wade et al.\ 2011b).  New {\em STIS} observations at
different phases -- including at magnetic equator edge-on ($\phi =
0.5$ and $|B_{\rm Z}| = 0$) -- have recently been obtained by our
group, and will elucidate the behavior of the P-Cygni profiles.

Another interesting issue raised by the new {\em HST} data is the
strength of the iron forest compared to previous {\em IUE}
(high-state) observations.  Improved agreement with the Fe$\;${\sc iv}
lines (and, surprisingly, with N$\;${\sc iv}~$\lambda$1718, He$\;${\sc
  ii}~$\lambda$1640, and other lines in the far~UV) was achieved by
adopting high turbulent velocities in the atmosphere models (up to 50
km s$^{-1}$). We speculate that shocks occurring about the magnetic
equator would be a possible explanation for the origin of such
high velocity dispersion in the inner regions (up to the Alfv\'en
radius); this would be consistent with the less intense lines seen at
other phases/viewing angles.  This hypothesis should be examined in
magnetic stars with physical properties similar to HD~108, and through
MHD simulations.

Overall, the emerging scenario for Of?p stars is that their
magnetic fields have a strong influence close to the photosphere, up
to the Alfv\'en radius.  Most of the observed spectral variability in
the optical is expected to arise from this region -- e.g., line
profiles changing between intense emission and absorption states, and
evidence for infall (see Paper~I). The moderate UV wind-line
variability (changes in $W_\lambda$ much less than found in the
optical) would reflect the fact that the extended stellar winds are
less affected by the magnetic field on a global scale (see Fig. \ref{LFRs}).

Notwithstanding that we have obtained a representative set of physical
parameters that generate models matching the spectrum of HD~108
reasonably well, a precise determination of the stellar and wind
parameters of Of?p stars awaits the construction of more sophisticated
atmosphere and MHD models. As yet, we are unable to model/predict the
spectrum of the gas which is strongly constrained by the magnetic
field. The discrepancies highlighted in the analysis carried out here,
and in our previous work (Paper~I) suggest that our atmosphere models
may be providing only rough estimates for the properties of these
objects.

Although we do not expect a large degree of asphericity in the wind of HD~108 
beyond the Alfv\'en radius -- given the observed UV variability and the reasonable fit achieved with
the current models (see Figures \ref{uvplot} and \ref{finalmodelfit})
-- multidimensional radiative transfer models may be needed to explain
the peculiar and variable optical and UV line profiles, in conformity
with constrained material subjected to rotational modulation. As shown by
MHD simulations, magnetically constrained winds may present a great
diversity of effects (ud-Doula et al.\ 2002, 2006, 2009). Common
assumptions, such as radiative equilibrium and a parametrized
distribution of X-ray emission, may need to be revised as well.

\section*{Acknowledgments}

WLFM acknowledges support from the Funda\c c\~ao de Amparo 
\`a Pesquisa do Estado do Rio de Janeiro (FAPERJ/APQ1). 
NRW acknowledges support provided by NASA through grant  
GO-12179.01 from STScI, which is operated 
by AURA, Inc., under NASA contract NAS5-26555.
YN acknowledges support from the Fonds National de la Recherche
Scientifique (Belgium), the Communaut\'e Fran\c cais de Belgique, the
PRODEX XMM and Integral contracts, and the 'Action de Recherche
Concert\'ee' (CFWB-Acad\'emie Wallonie Europe). GAW acknowledges 
Discovery Grant support from the Natural Sciences and Engineering 
Research Council of Canada (NSERC). AHD thanks support by the 
Spanish Ministerio de Ciencia e Innovaci\'on 
(grant AYA2010-21697-C05-04), the Consolider-Ingenio 2010 
Program (CSD2006-00070) and the Gobierno de Canarias 
(grant PID2010119). We thank the French Agence Nationale de la 
Recherche (ANR) for financial support. We also thank F. Martins, J. Sundqvist, 
and A. ud-Doula for valuable comments in an earlier version of 
the manuscript.

{}

\label{lastpage}

\begin{thebibliography}{}
\bibitem[]{} Barannikov, A. A., 2007, Information Bulletin on Variable Stars, 5756, 1
\bibitem[]{} Bouret, J.-C., Donati, J.-F., Martins F., Escolano C., Marcolino W., Lanz, T., Howarth, I. D., 2008, MNRAS, 389, 75
\bibitem[]{} Donati, J.-F., Babel, J., Harries, T. J., Howarth, I. D., Petit, P., Semel, M., 2002, MNRAS, 333,55
\bibitem[]{} Donati, J.-F., Howarth, I. D., Bouret, J.-C., Petit, P., Catala, C., Landstreet, J., 2006, MNRAS, 365, L6
\bibitem[]{} Grunhut, J. H., Wade, G. A., Marcolino, W. L. F., et al., 2009, MNRAS, 400L, 94
\bibitem[]{} Henrichs, H. F., Schnerr, R. S., ten Kulve, E., 2005, ASP Conference Series, 337, 114
\bibitem[]{} Hillier, D. J., Miller, D. L., 1998, ApJ, 496, 407
\bibitem[]{} Hillier, D. J., Lanz, T., Heap, S. R., Hubeny, I., Smith, L., Evans, C. J., Lennon, D. J., Bouret, J.-C., 2003, ApJ, 588, 1039
\bibitem[]{} Howarth, I. D., Walborn, N. R., Lennon, D. J., Puls, J., Naz\'e, Y., Annuk, K., et al., 2007, MNRAS, 381, 433
\bibitem[]{} Hubrig, S., Scholler, M., Schnerr, R. S., Gonz\'alez, J. F., Ignace, R., Henrichs, H. F., 2008, A\&A, 490, 793
\bibitem[]{} Hubrig, S., Scholler, M., Kharchenko, N. V., Langer, N., de Wit, W. J., Ilyin, I., Kholtygin, A. F., Piskunov, A. E., Przybilla, N., 2011, A\&A, 528, 151
\bibitem[]{} Kaper, L., Henrichs, H. F., Nichols, J. S., Snoek, L. C., Volten, H., Zwarthoed, G. A. A., 1996, A\&AS, 116, 257
\bibitem[]{} Marcolino, W. L. F., Bouret, J.-C., Martins, F., Hillier, D. J., Lanz, T., Escolano, C., 2009, A\&A, 498, 837
\bibitem[]{} Martins, F., Donati, J.-F., Marcolino, W. L. F., Bouret, J.-C., Wade, G. A., Escolano, C., Howarth, I. D., 2010, MNRAS, 407, 1423 (Paper~I)
\bibitem[]{} Naz\'e, Y., Vreux, J., Rauw, G., 2001, A\&A, 372, 195
\bibitem[]{} Naz\'e, Y., Rauw, G., Vreux, J., De Becker, M., 2004, A\&A, 417, 667
\bibitem[]{} Naz\'e, Y., ud-Doula, A., Spano, M., Rauw, G., De Becker, M., Walborn, N. R., 2010, A\&A, 520, 59 
\bibitem[]{} Shore, S. N., Brown, D. N., Sonneborn, G., Landstreet, J. D., Bohlender, D. A., 1990, ApJ, 348, 242
\bibitem[]{} Sim\'on-D\'iaz, S., Herrero, A., Uytterhoeven, K., Castro, N., Aerts, C., Puls, J., 2010, ApJL, 720, L174
\bibitem[]{} Smith, M. A., Fullerton, A. W., 2005, PASP, 117, 13
\bibitem[]{} Stahl, O., Kaufer, A., Rivinius, Th, et al., 1996, A\&A, 312, 539
\bibitem[]{} ud-Doula, A., 2008, Proceedings of the Workshop Clumping in Hot Star Winds, Potsdam, 125 
\bibitem[]{} ud-Doula, A., Owocki, S. P., 2002, ApJ, 576, 413
\bibitem[]{} ud-Doula, A., Owocki, S. P., Townsend, R. H. D., 2009, MNRAS, 392, 1022
\bibitem[]{} ud-Doula, A., Townsend, R. H. D., Owocki, S. P., 2006, ApJL, 640, L191
\bibitem[]{} Wade, G. A., et al., 2011, MNRAS, in press      
\bibitem[]{} Wade, G. A., et al., 2011b, MNRAS, 416, 3160    
\bibitem[]{} Walborn, N. R., 1972, AJ, 77, 312
\bibitem[]{} Walborn, N. R., Howarth, I. D., Herrero, A., Lennon, D. J., 2003, ApJ, 588, 1025
\bibitem[]{} Walborn, N. R., Sota, A., Ma\'iz Apell\'aniz, J., Alfaro, E. J., Morrell, N. I.,  Barb\'a, R. H., Arias, J. I., Gamen, R. C., 2010, ApJL, 711, L143
\end{thebibliography}
\end{document}